\documentclass{mn2e}
\usepackage{graphicx}

\title[Two-Phase Galaxy Formation]{Two-Phase Galaxy Formation}

\author[M. Cook et al.]{M. Cook$^{1,2}$, A. Lapi$^{3,1,4}$, G.L. Granato$^{4,2}$\\
$^{1}$Astrophysics Sector, SISSA/ISAS, Via Beirut 2-4, I-34014 Trieste, Italy\\
$^{2}$INAF, Osservatorio Astronomico di Padova, Vicolo dell' Osservatorio 5, I-35122 Padova, Italy\\
$^{3}$Dept. of Physics, Univ. di Roma `Tor Vergata', Via della Ricerca Scientifica 1, I-00133 Rome, Italy\\
$^{4}$INAF, Osservatorio Astronomico di Trieste, Via G.B. Tiepolo 11, I-34131 Trieste, Italy}

\begin{document}

\maketitle

\begin{abstract}
We propose and test a scenario for the assembly and evolution
of luminous matter in galaxies which substantially differs from
that adopted by other semianalytic models. As for the dark
matter (DM), we follow the detailed evolution of halos within
the canonical $\Lambda$CDM cosmology using standard Montecarlo
methods. However, when overlaying prescriptions for baryon
evolution, we take into account an effect pointed out in the
past few years by a number of studies mostly based on intensive
$N$-body simulations, namely that typical halo growth occurs in
two phases: an early, fast collapse phase featuring several
major merger events, followed by a late, quiescent accretion
onto the halo outskirts. We propose that the two modes of halo
growth drive two distinct modes for the evolution of baryonic
matter, favoring the development of the spheroidal and disc
components of galaxies, respectively. We test this idea using
the semianalytic technique. Our galaxy formation model
envisages an early coevolution of spheroids and the central
supermassive black holes, already tested in our previous works,
followed by a relatively quiescent growth of discs around the
preformed spheroids. In this exploratory study, we couple our
model to the spectrophotometric code \textsl{GRASIL}, and
compare our results on several properties of the local galaxy
population with observations, finding an encouraging agreement.
\end{abstract}


\section{Introduction}

A fundamental issue when modeling the evolution of galaxies in
a cosmological context is that the majority of the processes
driving baryonic evolution (such as star formation, various
feedback mechanisms, accretion onto supermassive black holes
[BHs]) operate or originate on scales well below the resolution
of any feasible simulation in a cosmic box. Moreover, these
processes are highly nonlinear, poorly understood from a
physical point of view, and approximated by means of
simplified, often phenomenological, and thus uncertain subgrid
prescriptions. Unfortunately, yet unsurprisingly, a number of
studies have clearly demonstrated that the results of these
models are heavily affected by different choices for such
prescriptions (e.g., Benson et al. 2003; Di Matteo et al.
2005), or for parameter values (e.g., Zavala et al. 2008).

Thus extensive comparisons between different scenarios and data
are generally conducted by means of semianalytic modeling
(SAMs) for baryons, often grafted onto gravity-only simulations
for the dark matter (DM) evolution. By definition of SAMs, the
general behavior of the system is outlined \emph{a priori}, and
then translated into a set of (somewhat) physically-grounded
analytical recipes --- suitable for numerical computation over
cosmological timescales --- for the processes which are
\emph{thought} to be more  relevant to galaxy formation and
evolution.

Although SAMs should not be viewed as complete first-principles
computations, they provide a convenient and powerful tool to
test an assumed galaxy formation scenario (i.e., the general
behavior and the adopted recipes) against existing data, and to
make predictions on future observations.

In general, SAMs (e.g., Cole et al. 2000; Hatton et al. 2003;
Cattaneo et al. 2005, 2006; Khochfar \& Silk 2006; Baugh et al.
2005; Bower et al. 2006; Croton et al. 2006; Monaco et al.
2007; Somerville et al. 2008), apart from relatively minor
variations, are constructed around two main assumptions: (i)
the initial outcome of gas cooling within DM halos is, at any
cosmic epoch, the development of a rotationally-supported disc
(since Rees \& Ostriker 1977; Silk 1977; White \& Rees 1978);
these discs usually undergo mild to moderate star formation
activity, unless extreme choices for the scaling of star
formation efficiency with galaxy properties are done; (ii) the
most natural driver of episodes of violent star formation at
any redshift is the merging of these gas-rich discs, which in
most models also constitutes the main channel for the formation
of spheroids, and in particular of large ellipticals (since
Cole 1991).

As a  result of this disc-merger-driven framework, baryons tend
to follow the hierarchical behavior of DM halos, and there is
no inherent relationship between the morphology and the star
formation history of galaxies. This is in sharp contrast with
the basic observational fact that low-mass galaxies tend to be
disc dominated, gas rich, blue, and actively star forming,
whilst more massive galaxies tend to be red, gas poor,
quiescent, and dominated by a spheroidal component mainly
comprised of old stars.

Due to these features, SAMs built around the two aforementioned
assumptions --- which from now on will be collectively referred
to as `standard SAMs' --- tend to be in tension with several
observations (e.g., Somerville et al. 2008), manifested by the
poor performances they had in anticipating observational
breakthroughs occurred more recently. For example, it is now
well established that baryonic structures undergo the
phenomenon referred to as `cosmic downsizing', whereby massive
star forming systems and associated supermassive BHs shined
mostly at high redshift, while smaller objects display
longer-lasting activity. Clearly, it is challenging to obtain
this behavior from the scheme outlined above; indeed, no model
did until relatively recently, and the present situation
remains unclear. In the past few years, almost all semianalytic
teams introduced simple recipes of feedback from active
galactic nuclei (AGNs) in their models, with the specific
target of quenching star formation in high mass galaxies at low
redshift. This additional ingredient significantly improves the
situation, but does not directly alleviate model difficulties
in producing enough massive systems at high $z$. As a result,
at least three state-of-the-art standard SAMs still do not
correctly reproduce the downsizing trend in stellar mass, nor
the {\it archeological downsizing} (Fontanot et al. 2009; see
below for more details).

A further example of challenges to models comes from the modest
evolution of the cosmic star formation activity above $z \sim
1$ (the so called Madau plot), strikingly at variance with
model predictions (e.g., Cole et al. 1994) generated before the
advent of surveys effective in discovering dust-enshrouded star
formation at high $z$ (Madau et al. 1996, 1998). It is fair to
note that a fraction, but not all, of the discrepancy was due
to the then adopted standard CDM cosmology and a lower
normalization for the fluctuation spectrum, resulting in
significantly more rapid evolution at high redshift than the
now favored $\Lambda$CDM.

In addition, even latest and most refined SAMs are seriously
challenged by the bright number counts and the high redshift
peak of $z$-distribution for submm galaxies. For instance,
Baugh et al. (2005) showed that the only way to reproduce the
statistic of submm sources, usually considered the precursor of
local ellipticals, in the context of their standard SAM, is to
adopt an extremely top-heavy intial mass function (IMF) during
galaxy-merger-induced starbursts. However, their model predicts
masses of submm sources likely too low by more than one order
of magnitude (Swinbank et al. 2008), and still shows
discrepancies with observed trends of $\alpha/$Fe in local
ellipticals (Nagashima et al. 2005).

Without doubt, the field of galaxy formation is \emph{led} by
observations. Indeed, physical processes have been continuously
added to SAMs, or existing ones have been substantially revised
by SAM developers in order to face serious mismatches between
model outputs and new datasets. Besides many relatively minor
but subtle details, major examples comprise a treatment of the
growth of supermassive BHs in galaxy centers and of the ensuing
energetic feedback from nuclear activity (Granato et al. 2004
[G04]\footnote{in the context of a nonstandard SAM focused on
the coevolution of quasars and spheroids, of which this paper
can be considered an extension, see below.}; Bower et al. 2006;
Croton et al. 2006; Monaco et al. 2007; Somerville et al.
2008)\footnote{see also Hatton et al. (2003) and Cattaneo et
al. (2007) for the effect of an highly idealized criterium of
`pseudo AGN feedback', phenomenologically inspired by the
Magorrian (1998) relationship.}, the effects of `cold' versus
`hot' accretion flows onto DM halos, as suggested by Dekel \&
Birnboim (2006) and implemented in a full SAM by Cattaneo et
al. (2006; see also Somerville et al. 2008), or an extremely
top-heavy flat IMF in merger-driven bursts (Baugh et al. 2005).
These examples show that the complexity and degrees of freedom
of standard SAMs have been steadily increased by modelers in
order to improve the agreement with the data, but despite these
efforts several points of tension still remain (see Monaco et
al. 2007).

Within this paper, we follow a significantly different approach
and submit a novel scenario for galaxy formation, modifying the
very basic assumptions of standard SAMs that, according to us,
are the origin of their tensions with observations. Our
scenario envisages that the fundamental dichotomy between the
spheroid and disc components in galaxies reflects two
dominating modes for the assembly of visible matter, feasibly
being ultimately driven by two dominating modes governing the
growth of DM halos (see below). For typical $L^*$ galaxy halos,
$z\ga 2$ corresponds to an era dominated by violent merging
episodes, leading to huge bursts of star formation and to the
observed coevolution of spheroids with hosted central
supermassive BHs; $z\la 2$ instead correspond to an era where
the most relevant process is quiescent accretion of matter
yielding, under suitable conditions, the formation of discs
around preexisting spheroids.

Many studies in the literature on the chemical and
spectrophotometric properties of local galaxy populations
(\emph{stellar archeology}; see Thomas et al. 2005 and
references therein; Chiappini et al. 1997; Portinari \& Chiosi
1999) reached the broad conclusion that galaxies of later type,
which are less massive on average, formed their stars at later
times and over a longer period (see also Gavazzi et al. 1996).
This phenomenon is sometimes referred to as
{\emph{archeological downsizing} and is not reproduced by three
state-of-the-art SAMs (Fontanot et al. 2009; however, see
Kaviraj et al. 2005 for a discussion of color-magnitude
relation in cluster ellipticals as a test for hierarchical
models). These conclusions have more recently been confirmed by
modern surveys at high redshift, directly showing that the
sites of active star formation shift from high-mass galaxies at
early times to lower-mass systems at later times
(\emph{downsizing in time}; Cowie et al. 1996; Guzman et al.
1997; Brinchmann \& Ellis 2000; Kodama et al. 2004; Juneau et
al. 2005; Bell et al. 2005; Noeske et al. 2007). Further
support for two different epochs and formation mechanisms of
spheroids and discs comes from the analysis of the colour and
structural properties of decomposed galaxy components in the
Millennium Galaxy Catalog (Driver et al. 2006).

\begin{figure}
\centering
\includegraphics[type=eps,ext=.eps,read=.eps, height=6.5cm, angle=0]{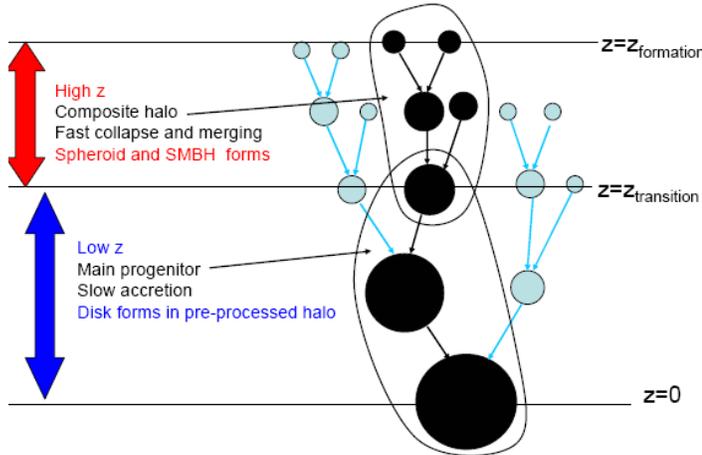}\\
\caption{A schematic of the two-phase evolution of DM halos and
associated modes of galaxy formation.}
\end{figure}

From the theoretical perspective, recent analysis of high
resolution simulations of individual DM halos forming in
cosmological volumes (Zhao et al. 2003; Diemand et al. 2007;
Hoffman et al. 2007; Ascasibar \& Gottloeber 2008) have
provided support for our picture. In these studies, two
distinct phases in the growth of DM halos have been identified:
an early fast collapse featuring a few violent major mergers,
and a later calmer phase including many minor mergers and
smooth accretion. During the early collapse a substantial mass
is gathered through major mergers, which effectively
reconfigure the gravitational potential wells and cause the
collisionless DM particles to undergo dynamical relaxation and
isotropization (Lapi \& Cavaliere 2009); therefrom the system
emerges with a definite structure for the inner density and
gravitational potential (Lu et al. 2006). During the later
quiescent phase, moderate amounts of mass are slowly accreted
mainly onto the halo outskirts, little affecting the inner
structure and potential, but quiescently rescaling the overall
mass upwards. Mo \& Mao (2004) qualitatively suggested that
this two-phase formation of DM halos may be at the origin of
the disc-spheroid dichotomy, alleviating several problems of
the standard SAMs. Here we take up this general idea, and
construct a full semianalytic model capable at making
quantitative predictions to be compared with present and future
observations. We explicitly point out that the backbone for the
cosmological growth of DM halos is broadly the same as that
adopted by all other SAMs. The novelty is in the fact that we
identify the transition between the two phases, and we assume
that the main processes driving the evolution of luminous
matter are strongly linked to the two different modes of DM
assembly.

Our model constitutes a natural extension to include disc
formation at low-$z$, of the Antihierarchical Baryon Collapse
scenario (\textsl{ABC}; G04; Granato et al. 2006; Silva et al.
2005; Lapi et al. 2006; Mao et al. 2007; Lapi et al. 2008) that
was focused on the high redshift coevolution between spheroidal
galaxies and hosted supermassive BHs. This has been extremely
successful in reproducing a wealth of observations, including
statistics of submm galaxies, properties of local elliptical
galaxies, the results of deep $K$-band surveys, the estimated
local mass function of supermassive BH, and statistics of
high-redshift quasars (QSOs). These results are essentially
inherited by the model presented here.

In summary, motivated by the successes of the high redshift
\textsl{ABC} framework and the compelling theoretical and
observational evidence in support of a two-phase galaxy
formation scenario, we have developed a framework linking high
and low redshift processes in order to generate the observed
dichotomy between early-type and late-type galaxies. The plan
of the paper is the following: we describe in detail the
two-phase DM treatment in \S~2; in \S~3 we describe the
modeling of the baryonic matter evolution for the spheroid and
disc components; we present our results in \S~4 and finally we
summarize and discuss our findings in \S~5.

Throughout the paper we adopt the standard $\Lambda$CDM
concordance cosmology, as constrained by \textsl{WMAP} 5-year
data (Spergel et al. 2007). Specifically, we adopt a flat
cosmology with density parameters $\Omega_M=0.27$ and
$\Omega_{\Lambda}=0.73$, and a Hubble constant $H_0=70$ km
s$^{-1}$ Mpc$^{-1}$.

\section{Dark Matter Sector}

In this work we compute the mass growth histories of DM halos
using a binary mergertree with accretion based on the extended
Press \& Schechter formalism (see Lacey \& Cole 1993);
specifically, we rely on the algorithm originally developed by
Cole et al. (2000), and recently improved by Parkinson et al.
(2008) to reproduce the outcomes of $N$-body simulations.

The algorithm starts from the expression for the mass fraction
of a halo with mass $M_2$ at redshift $z_2$ that was contained
within a progenitor halo of mass $M_1<M_2$ at $z_1>z_2$:
\begin{equation}
f = \frac{\Delta
\delta_{c}}{\sqrt{2\pi\,(\Delta\sigma^{2})^{3}}}\,
e^{-(\Delta\delta_{c})^2/2\, \Delta\sigma^2}\,
\left|\frac{\mathrm{d}\sigma^{2}}{\mathrm{d}M}\right|_{M=M_1}~;
\end{equation}
here $\Delta\delta_c = \delta_c(z_1)-\delta_c(z_2)$ is the
difference between the linear thresholds for collapse at
redshifts $z_1$ and $z_2$, while $\Delta\sigma^2 =
\sigma^2(M_1)-\sigma^2(M_2)$ is the difference between the
variances of the linear density fluctuations extrapolated at
$z=0$ in spheres containing masses $M_1$ and $M_2$.

Taking the limit of the above equation as $z_1 \rightarrow z_2$
one finds the merger rate as
\begin{equation}
\frac{\mathrm{d}f}{\mathrm{d}z} =
\frac{1}{\sqrt{2\pi\,(\Delta\sigma^{2})^{3}}}\,
\left|\frac{\mathrm{d}\delta_{c}}{\mathrm{d}z}\right|\,
\left|\frac{\mathrm{d}\sigma^{2}}{\mathrm{d}M}\right|_{M=M_1}~;
\end{equation}
now one easily can work out the distribution for the number of
halos with mass $M_{1}$ into which a halo with mass $M_2$
splits during a step $\mathrm{d}z$ up in redshift:
\begin{equation}
\frac{\mathrm{d}N}{\mathrm{d}M_{1}} =
\frac{\mathrm{d}f}{\mathrm{d}z}\, \frac{M_{2}}{M_{1}}\,
\mathrm{d}z~.
\end{equation}

Then given a mass resolution $M_{\mathrm{res}}$ one may define
the mean number of progenitors with masses $M_1$ between
$M_{\mathrm{res}}$ and $M_2/2$:
\begin{equation}
P = \int_{M_{\mathrm{res}}}^{M_{2}/2}
\frac{\mathrm{d}N}{\mathrm{d}M_{1}}\,\mathrm{d}M_{1}~,
\end{equation}
and the fraction of mass of the final object in progenitors
below the resolution limit:
\begin{equation}
F = \int_{0}^{M_{\mathrm{res}}}
\frac{\mathrm{d}N}{\mathrm{d}M_{1}}\, \frac{M_{1}}{M_{2}}\,
\mathrm{d}M_{1}~.
\end{equation}
In fact, in constructing the above quantities we have replaced
\begin{equation}
\frac{\mathrm{d}N}{\mathrm{d}M_{1}} \longrightarrow
\frac{\mathrm{d}N}{\mathrm{d}M_{1}}\, G(\sigma_{1}/\sigma_{2},
\delta_{2}/\sigma_{2})
\end{equation}
where $G(\sigma_{1}/\sigma_{2}, \delta_{2}/\sigma_{2})$ is a
perturbing function given by Parkinson et al. (2008), tuned to
bring the Montecarlo merger histories in close agreement with
the outcomes of state-of-the-art numerical simulations.

Given all that, the mergertree algorithm is straightforward:
choose a mass $M_2$ and redshift $z$ for the final halo; pick
up a step in redshift $\mathrm{d}z$ such that $P\ll 1$ to
ensure that the halo is unlikely to have more than two
progenitors at $z+\mathrm{d}z$; generate a uniform random
number $R$ between 0 and 1; if $R>P$ do not split the main halo
at this step, and simply reduce its mass to $M_2\,(1-F)$ to
account for sub-resolution accretion; if $R\leq P$ generate a
random value of $M_1$ between $M_{\mathrm{res}}$ and $M_2/2$
consistent with the distribution given by Eq.~(1), to produce
two new halos with masses $M_1$ and $M_2\,(1-F)-M_1$; repeat
the process on each new halo at successive redshift steps to
build the overall merging tree.

In our implementation of the algorithm we use an adaptive step
size $\Delta z$ such that $P$ remains at a value significantly
below unity, and then postprocess the tree by sampling it over
convenient redshift intervals. In addition, we take the
resolution mass $M_{\mathrm{res}}(z)$ as the one corresponding
to a virial temperature of $10^{4}\,K$, above which atomic gas
cooling allows baryonic structures to condense.

As an input of the algorithm, we use the Bardeen et al. (1986)
power spectrum of density fluctuations with correction for
baryons (Sugiyama 1995), normalized so as to yield a mass
variance $\sigma_8\approx 0.8$ on a scale of $8\, h^{-1}$ Mpc.
As an output, we obtain many realizations (up to several
thousands within conceivable computational times) of the
mergertree for a given present mass $M_0$ at $z=0$; each
realization lists all the progenitors of $M_0$ at different
redshifts and describes how and when these merge together. We
generate trees for masses $M_0$ spanning the range from
$10^{9}$ to $10^{14}\,M_{\odot}$ in logarithmic increments, and
follow the related merging histories down to the resolution
mass.

We find that the halo growth along a given evolutionary track
occurs in two distinct phases: an early violent collapse where
rapid growth is enforced by major mergers among several massive
clumps; and a late period of gentle mass addition through
calmer accretion (see Fig.~1) extending down to the present
time. A similar behavior has been pointed out in a number of
recent numerical simulations (Weschler et al. 2002; Zhao et al.
2003; Diemand, et al. 2007; Hoffman et al. 2007; Ascasibar \&
Gottloeber 2008), and has been analyzed in semianalytic studies
of Montecarlo merging trees to explore the origin of the
structural properties of DM halos (see Lu et al. 2006; Li et
al. 2007; Lapi \& Cavaliere 2009).

In our view these different evolutionary phases of DM halo
growth should significantly affect the main physical processes
regulating the evolution of the baryonic matter within them; in
particular, we envisage the violent early collapse phase to be
associated with the formation of the spheroid and hosted
supermassive BH, while the gentle late phase to be favorable
for the stable growth of galaxy disc around the preexisting
spheroid-BH structure.

As to the halo spatial structure we assume the standard NFW
(Navarro, Frenk \& White 1997) density profile
\begin{equation}
{\rho(r)\over \rho_c} = {\Delta_{\rm vir}\,c^2\,g(c)\over
3\,(r/r_{\rm vir})\,(1+c\,r/r_{\rm vir})^2}~;
\end{equation}
here $\rho_c$ is the critical density, $g(c)\equiv
[\log(1+c)-c/(1+c)]^{-1}$ is a weak function of the
`concentration' parameter $c$, and $\Delta_{\rm vir}\approx
18\pi^2+82\,[\Omega_M(z)-1]-39\, [\Omega_M(z)-1]^2$ is the
non-linear collapse threshold in terms of the evolved matter
density parameter
$\Omega_M(z)=\Omega_M\,(1+z)^3/[\Omega_M\,(1+z)^3+\Omega_\Lambda]$.
In fact, $N$-body experiments (Taylor \& Navarro 2001; Zhao et
al. 2003; Diemand et al. 2007; Hoffmann et al. 2007; Ascasibar
\& Gootloeber 2008) show that the NFW profile is established
during the early fast collapse phase, with concentration
parameter $c_t \simeq 4$ for $z\ga z_t$; in the slow accretion
phase for $z\la z_t$ the DM halo potential well retains its
shape and $r_s$ stays put while the overall size $r_{\rm vir}$
of the system increases, to the effect of rising the
concentration parameter, see below.

Equipped with these notions, we effectively trace the redshift
evolution in the tree of a DM halo mass $M_z$ with present
value $M_0$ as follows (see Fig.~1). First of all, we compute
the concentration $c_0$ of the mass $M_0$ according to the
prescription by Macci\`o et al. (2007):
\begin{equation}
\log c_0 = 1.071 - 0.098\, \left[\log \left({M_0\over
M_{\odot}}\right)-12\right]~;
\end{equation}
we stress that our computation neglects the scatter of $c_0$ at
fixed mass, and any dependence of $c_0$ itself on the details
of the merging history (see Wechsler et al. 2002; Zhao et al.
2003). In addition, for the sake of simplicity we disregard the
influence of the baryons on the halo structure in terms of
adiabatic contractions or expansions (see Blumenthal et al.
1986; Gnedin et al. 2004).

Then during the late, slow accretion phase we take $M_z$ as the
mass of the main progenitor, i.e., we follow only the main
branch of the mergertree. This is achieved by starting at $z=0$
and working toward higher redshifts, taking the most massive
halo at each splitting (merger) event. We neglect the baryonic
processes (in particular star formation) occurring in the other
branches of the tree; in other words, we make the approximation
that when matter in the minor branches of the tree joins the
main progenitor it is still pristine, unevolved by baryonic
processes, and thus mainly in the form of a gaseous medium with
primordial metallicity. This is conceivable since: these halos
have masses close to the cooling mass, and thus have just grown
from below this threshold; we expect the star formation
efficiency to be relatively low there, since supernova (SN)
feedback would be most effective in ejecting gas from such
shallow potential wells; such small halos are likely to contain
small galaxies that would not alter significantly the
properties of the galaxy hosted by the main progenitor if they
happened to merge with it.

\begin{figure}
\centering
\includegraphics[type=eps,ext=.eps,read=.eps, height=8.5cm, angle=90]{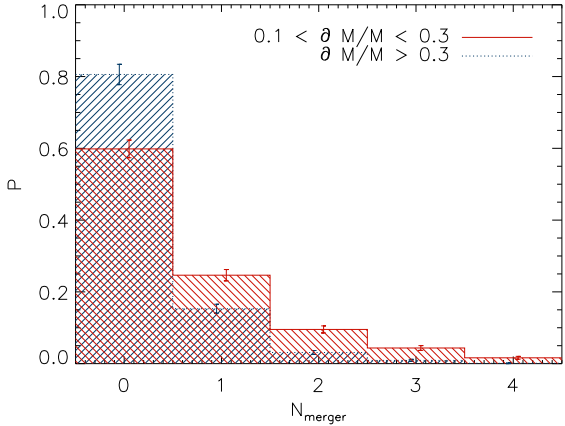}
\caption{Probability distribution for the number of mergers
undergone by DM halos during the slow accretion phase (results
over $1000$ realizations); various linestyles refer to
different ranges for the relative mass addition $\Delta M/M$.
In general, major mergers between halos are very rare events
during the slow phase.}
\end{figure}

\begin{figure}
\centering
\includegraphics[type=eps,ext=.eps,read=.eps, height=8.5cm, angle=90]{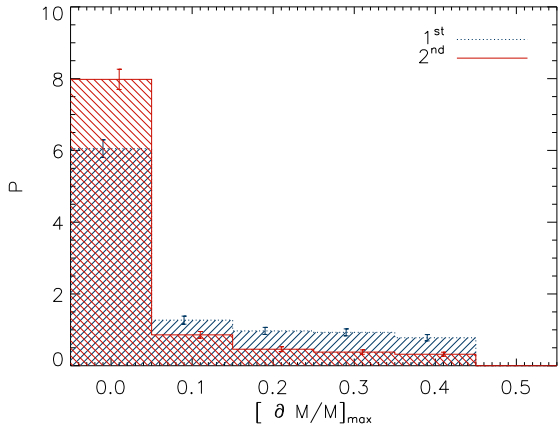}
\caption{Probability distribution for the relative mass
addition to halos during the slow accretion phase (results over
$1000$ realizations); various linestyles refer to the largest
and the second largest value of $\Delta M/M$ in the tree during
the slow accretion phase. In general the dominant mechanism for
growing halos in the slow phase is the accretion of small
lumps.}
\end{figure}

\begin{figure}
\centering
\includegraphics[type=eps,ext=.eps,read=.eps, height=8.5cm, angle=90]{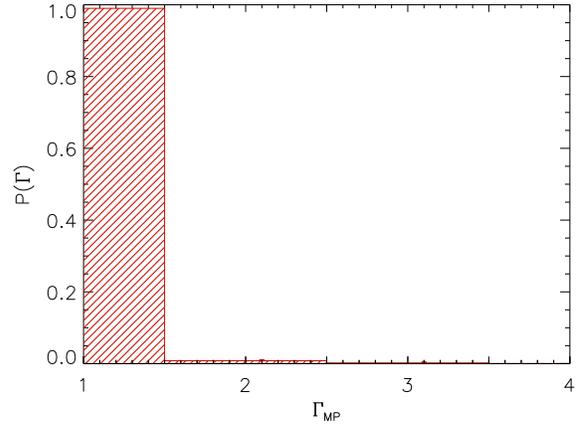}
\caption{Probability distribution for the mass rank of the main
progenitor in the mergertree at $z_t$. In general, the main
progenitor is actually the most massive halo at the
transition.}
\end{figure}

\begin{figure}
\centering
\includegraphics[type=eps,ext=.eps,read=.eps, height=8.5cm, angle=90]{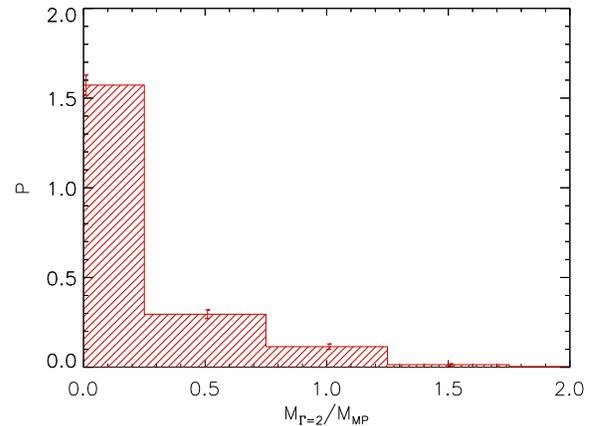}
\caption{Probability distribution for the mass ratio between
the main progenitor and the most massive of all other halos in
the mergertree at $z_t$; in general, all the other halos are
significantly less massive than the main progenitor.}
\end{figure}

\begin{figure}
\centering
\includegraphics[type=eps,ext=.eps,read=.eps, height=8.5cm, angle=90]{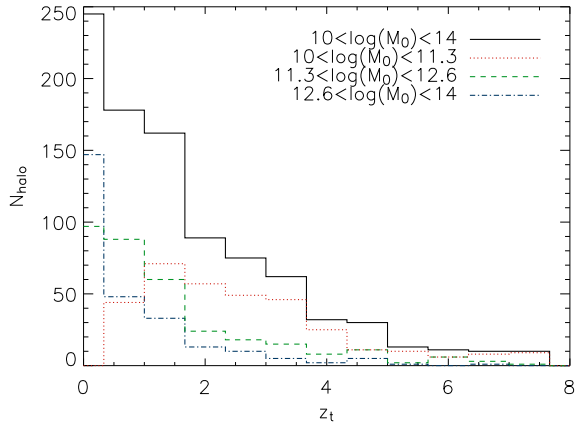}
\caption{Probability distribution for the transition redshifts
(results over $1000$ realizations); various linestyles are for
different ranges of present halo mass $M_0$. The distributions
of $z_t$ are roughly lognormal with a low redshift peak and a
rather extended tail at high $z$. Note that lower mass halos
have a higher $z_t$ on the average.}
\end{figure}

\begin{figure}
\centering
\includegraphics[type=eps,ext=.eps,read=.eps, height=8.5cm, angle=90]{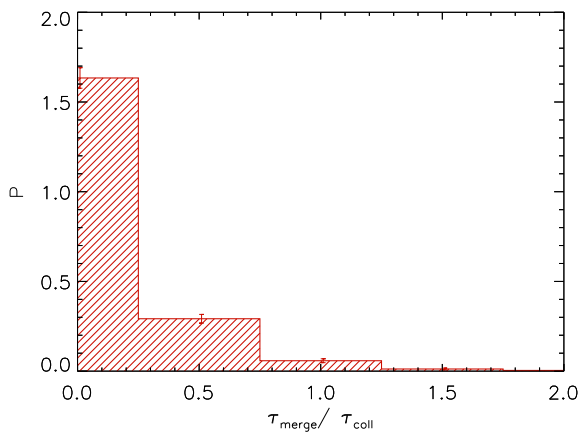}
\caption{Probability distribution for the ratio between the
merging time and the dynamical time of the individual merging
units during the early, fast collapse phase (results over
$1000$ realizations). In general, during the early collapse
mergers occur on very short timescales.}
\end{figure}

\begin{figure}
\centering
\includegraphics[type=eps,ext=.eps,read=.eps, height=8.5cm, angle=90]{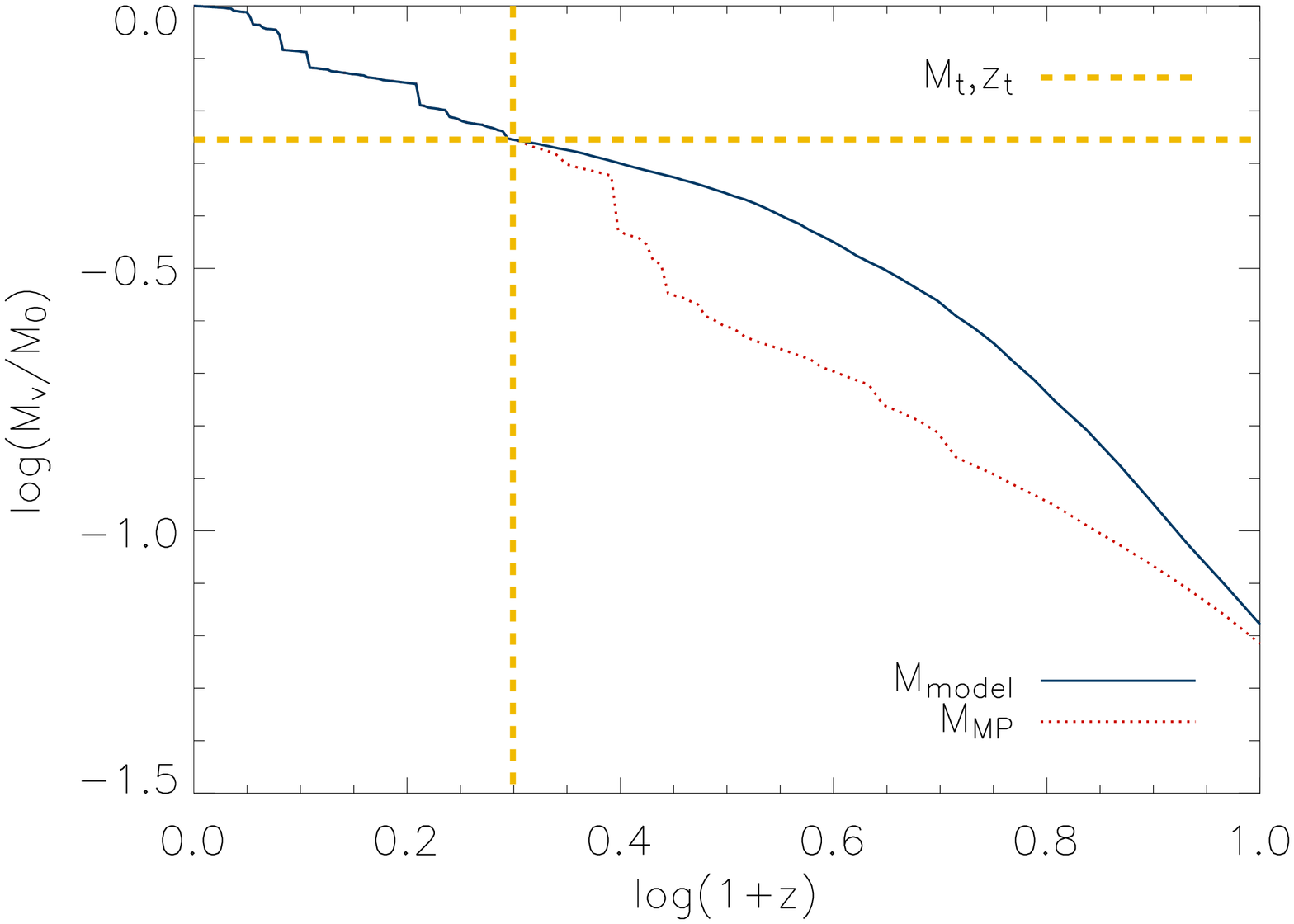}
\caption{Typical mass accretion history for a Milky Way-sized
DM halo ($M_0\approx 5\times 10^{11}\, M_{\odot}$). The transition
between slow accretion and fast collapse, that occurs at a
redshift $z_t\approx 1$, is highlighted as a dashed line. The
halo growth (solid line) follows the main
progenitor's one (dot-dashed line) during the late slow accretion
phase, and the composite halo (see text for details)
during the early fast collapse.}
\end{figure}

For a quantitative analysis, we compute and illustrate in
Figs.~2 and 3 the properties of mergers undergone by DM halos
along their growth history during the slow accretion phase. We
find that DM halos grow predominantly through small accretion
events and that the majority of our halos (about $80\%$) do not
undergo a substantial merger event; we recall that
conventionally a major merger is defined as one in which the
added mass exceeds that of the merging units by $1/3$ or more,
i.e. $\Delta M/M > 1/3$, see Lacey \& Cole (1993). Note,
however, that a minority of halos ($20\%$ or so) do undergo a
major merger event; plainly, within these systems the growth of
stable galaxy discs can be temporarily interrupted. In
addition, we note that the number of major mergers during the
slow accretion phase is closely independent of halo mass; thus
the average major merger rate is higher for more massive halos,
which have more recently made their transition into the slow
accretion phase.

We compute the concentration $c_z$ associated to the mass $M_z$
after Zhao et al. (2003) using
\begin{equation}
\frac{[\log(1+c_z)-c_z/(1+c_z)]\,c_{z}^{-3\alpha}}
{[\log(1+c_0)-c_0/(1+c_0)]\,c_0^{-3\alpha}} =
\left(\frac{H_z}{H_0}\right)^{2\alpha}
\left(\frac{M_z}{M_0}\right)^{1-\alpha}~;
\end{equation}
here $\alpha=0.48$ ($0.64$) in the slow (fast) phase is a
fitting parameter derived from $N$-body simulations, and
$H_z=H_0\,[\Omega_M\,(1+z)^3+\Omega_\Lambda]^{1/2}$  is the
evolved `Hubble constant'; a less accurate but simpler
approximation of the expression above in the slow phase reads
$c_z=c_0\,[H_0/H_z]^{1/\eta}$ with $\eta\sim 1$, see Mo \& Mao
(2004). Following Zhao et al. (2003), the transition between
the slow accretion and the fast major merger phase occurs at
the redshift $z_t$ where the concentration $c_z$ decrease below
the critical value $c_t=4$.

In Figs.~4 and 5 we illustrate the properties of the mergertree
at $z_t$. First, we note that the main progenitor is generally
the most massive halo in the mergertree at $z_t$; second, the
ratio between the mass of the main progenitor halo to the
second most massive halo is generally very large. These
findings show that, during the late slow accretion phase, the
relevant evolution of the main progenitor is characterized by
minor merger events with low mass halos.

It is interesting to note (see Fig.~6) that on average the
transition redshift $z_t$ decreases with increasing present day
halo mass. In our scenario this means that larger halos have
less or no time to develop a substantial disc component, hence
massive halos will tend to host pure spheroids and viceversa,
in broad agreement with observations. Incidentally, note that
although massive halos spend the majority (or all) of their
lifetimes within the fast collapse phase, the timescales for
baryonic evolution, that are governed by the coevolution of the
spheroid and the central supermassive BH, can be much shorter;
in fact, these systems typically become `red and dead' due to
QSO feedback at relatively high redshifts, and passively evolve
thereafter until the present (see \S~3.1).

At $z > z_t$, during the fast collapse phase, we compute $M_z$
not as the main progenitor mass, but as that of the composite
halo made of the overall mass in all the branches of the tree
that will contribute to the main progenitor mass at $z_t$.
Moreover, we consider only halos whose mass exceeds the
critical halo mass for efficient gas cooling, i.e., we consider
only halos where the related virial temperature is above $10^4$
K.

Quantitatively, this can be justified by comparing the
dynamical time for the composite halo with the major merger
timescale. The former is defined as the timescale for the
material to ballistically collapse to its center:
\begin{equation}
\tau_{\mathrm{coll}} = \left(\frac{3\pi}{32\,G\,\bar{\rho}} \right)^{1/2}~,
\end{equation}
in terms of the average density of the composite halo; the
latter is defined as the average timescale for a major merger
throughout the duration of the fast collapse phase:
\begin{equation}
\tau_{\mathrm{merge}} = \frac{N_{MM}}{\delta t_{\rm fast}}~,
\end{equation}
where $N_{MM}$ is the number of times the main progenitor
undergoes a major merger during the fast collapse phase, and
$\delta t_{\rm fast}$ is the timelapse the halo spends within
it.

Fig.~7 shows that during the fast collapse phase several major
merger events occur over timescales typically shorter than the
dynamical time of the composite system, so just following the
composite halo is a conceivable approximation.

In Fig.~8 we plot one typical realization of the mass accretion
history for a Milky Way-sized DM halo, highlighting the
transition redshift $z_t$, the behavior of the composite halo
at $z>z_t$ and of the main progenitor at $z<z_t$.

Finally, we specify the angular momentum of the DM halo in the
mergertree as follows. First of all, we recall that the angular
momentum $J$ is usually expressed in terms of the dimensionless
spin parameter $\lambda \equiv J\,|E|^{1/2}\, G^{-1}\,
M^{-5/2}$, where $E$ is the total energy of the halo; $N$-body
experiments have shown that $\lambda$ does not correlate with
halo mass nor concentration, is nearly independent of the
redshift, and follows a lognormal distribution with average
value $0.04$ and scatter $0.5$ dex (Bullock et al. 2001;
Macci\`o et al. 2007).

Thus for each halo we randomly select a value of $\lambda$ from
such a distribution and neglect its evolution (e.g., Barnes \&
Efstathiou 1987; Kravstov et al. 1997; Vitvitska et al. 2002;
Hetznecker \& Burkert 2006). We are aware that the latter could
impact on the evolution of disc properties, but choose to keep
our treatment of this effect as simple as possible, given that
to our knowledge a robust modeling has not yet been included
into state-of-the-art SAMs; e.g., Somerville et al. (2008) use
the spin parameter of the more massive halo at any given merger
event.

\section{Baryonic sector}

\subsection{Modeling the Spheroid}

The treatment of the baryonic processes in the fast major
merger phase follows the recipes adopted by G04 to model the
coevolution of spheroids and supermassive BHs, and already
exploited by our team in several previous papers (Silva et al.
2005; Cirasuolo et al. 2005; Lapi et al. 2006; Granato et al.
2006; Mao et al. 2007; Lapi et al. 2008). Here we provide a
qualitative summary of the model focusing on its distinctive
features, and defer the reader to G04 for all the details.

We recall from \S~2 that during the fast collapse phase, a
rapid sequence of major mergers build up a DM halo of mass
$M_t$ at the transition redshift $z_t$; as for baryonic matter,
we assume that condensation and cooling processes become
effective at a formation redshift $z_f > z_t$ when the mass of
the composite halo surpasses a substantial fraction of $M_t$,
namely $M_t/2$ as widely adopted in the literature to define
the `formation epoch' (Lacey \& Cole 1993; Kitayama \& Suto
1996). The results presented in this paper are almost
insensitive to the exact choice for the fraction of $M_t$
adopted in the definition of $z_f$. Specifically, we have
checked that galaxy properties vary less than $10\%$ when the
mass fraction is changed between approximately $1/5$ and $2/3$,
due to the strong effects of QSO feedback in terminating the
star formation soon after $z_f$.

After $z_f$, a mass $M_{\rm inf}\approx f_b\,M_t$ of baryonic
matter, in cosmic proportion $f_b\approx 0.17$ with the DM's,
is shock heated to the virial temperature by falling into the
gravitational potential well. This hot gas, assumed to follow
an isothermal distribution, may cool quickly especially in the
denser central regions at the rate
\begin{equation}
\dot{M}_{\rm cool}={M_{\rm inf}\over t_{\rm cool}}~,
\end{equation}
in terms of the local cooling timescale
\begin{equation}
t_{\rm cool}={3\,\rho_{\rm gas}\,kT\over 2\mu m_p\,\mathcal{C}\,
n_e^2(r)\, \Lambda(T)}~;
\end{equation}
here $\rho_{\rm gas}$ is the gas density, $n_e$ is the electron
density, $T$ is the temperature, $\Lambda(T)$ is the cooling
function, and $\mathcal{C}\sim 5-10$ is a parameter describing
the clumpiness of the gas.

The cooled gas mass $M_{\rm cold}$, assumed to still follow the
DM radial distribution, may form stars directly over the local
dynamical timescale, providing a rate of star formation
\begin{equation}
\psi(t) = \int_0 ^{r_{\rm vir}} \frac{1}{t_{\rm dyn}(r)}\,
\frac{{\mathrm d}M_{\rm cold}(r,
t)}{{\mathrm d}r}\,{\mathrm d}r~,
\end{equation}
with
\begin{equation}
t_{\rm dyn}=\left[{3\pi\over 32\,G\rho(r)}\right]^{1/2}~.
\end{equation}
This is conceivable during this evolutionary stage since the
ongoing major mergers continuously reshuffle the gravitational
potential, enforcing dynamical relaxation and orbit
isotropization of the collisionless DM and stellar components
(Lapi \& Cavaliere 2009). We recall that usually SAMs assume
instead that the first result of gas cooling is the formation
of rotationally supported discs, characterized by much milder
star formation activity, since the adopted star formation
timescale is typically much longer than some dynamical
times\footnote{for instance, $50\, t_{\rm dyn}$ in Hatton et
al. (2003), $\sim 200\, t_{\rm dyn}$ in Cole et al. (2000),
$\sim 350\, t_{\rm dyn}$ in Bower et al. (2006), and $\sim 15\,
t_{\rm dyn}$ in Croton et al. (2006).}.

With our prescriptions, large galactic halos can attain star
formation rates of the order of $\sim 1000$ solar masses per
year over timescales of a fraction of Gyr. This is required to
explain the submm galaxy population without invoking an
extremely top-heavy IMF (e.g., Baugh et al. 2005). In fact, our
IMF has the standard Salpeter slope $1.25$ in the high mass
tail, and flattens to a slope $0.4$ below $1\, M_{\odot}$. As
shown in Romano et al. (2005), this performs better than the
Salpeter one in reproducing the detailed chemical properties of
elliptical galaxies.

Star formation promotes the gathering of some cool gas into a
low-angular-momentum reservoir around the central supermassive
BH. A viable mechanism for this process is radiation drag (see
discussion by Umemura 2001; Kawakatu \& Umemura 2002; Kawakatu,
Umemura \& Mori 2003), which has the nice feature of predicting
a mass transfer rate to the reservoir proportional to the SFR
to a good approximation:
\begin{equation}
\dot{M}_{\rm RD}= \alpha_{\rm RD}\times 10^{-3}\,(1-e^{-\tau_{\rm RD}})\,\psi(t)~.
\end{equation}
The constant of proportionality $\alpha_{\rm RD}\sim 1-5$ can
be fixed to produce a good match to the correlation between the
spheroid and the supermassive BH masses observed in the local
universe. The quantity
\begin{equation}
\tau_{\rm RD}\approx \tau_{0}\,\left({Z\over Z_{\odot}}\right)\,
\left({M_{\rm cold}\over 10^{12}\,M_{\odot}}\right)^{1/3}
\end{equation}
represents the effective optical depth of the gas clouds in
terms of the normalization parameter $\tau_0\sim 1-5$; for more
details, see the discussion around Eqs.~(14) to (17) in G04.

Eventually, this gas accretes onto the BH powering the nuclear
activity; in this early phase plenty of material is supplied to
the BH, so that the latter can accrete close to the Eddington
limit
\begin{equation}
\dot{M}_{\rm BH}= \lambda_{\rm Edd}\,{1-\eta\over \eta}{M_{\rm BH}\over t_{\rm Edd}}~,
\end{equation}
and grows almost exponentially from a seed of $10^2\,
M_{\odot}$. The $e$-folding time involves the Eddington time
$t_{\rm Edd}\approx 4\times 10^8$ yr, the radiative efficiency
$\eta\sim 0.15$, and the actual Eddington ratio $\lambda_{\rm
Edd}\sim 0.3-3$.

The energy fed back to the gas by SN explosions and BH activity
regulates the ongoing star formation and BH growth. The two
feedback processes have very different dependencies on halo
mass and on galaxy age (e.g., on the time since $z_f$). The
feedback due to SN explosions removes the starforming gas at a
rate
\begin{equation}
\dot{M}_{\rm SN} = -\frac{2}{3}\, \epsilon_{\rm
SN}\, \frac{\eta_{\rm SN}\, E_{\rm SN}}{\sigma^2}\, \psi(t)~;
\end{equation}
here $\sigma^2$ is the velocity dispersion within the bulge,
$E_{\rm SN}\approx 10^{51}$ erg is the energy released in a
single SN event, $\eta_{\rm SN}$ is the number of Type II SNe
expected per solar mass of formed stars (determined by the
IMF), and $\epsilon_{\rm SN}\sim 0.05$ is the fraction of this
energy which is effectively coupled to the gas. Thus the SN
feedback evolves almost in parallel with the star formation; it
is very effective in low-mass halos severely limiting the
growth of stellar and BH components there, but is of minor
importance in the more massive galactic halos.

The QSO feedback considered by G04 acts both on the cold as
well as on the hot gas, unbinding them from the DM halo
potential well at a rate
\begin{equation}
\dot{M}_{\rm QSO}\simeq -2 \times 10^{3}\, \epsilon_{\rm
QSO}\, \frac{L_{{\rm Edd}, 46}^{3/2}}{(\sigma/300\,
\mathrm{km}\,\mathrm{s}^{-1})^2}\, M_{\odot}\,\mathrm{yr}^{-1}~;
\end{equation}
this functional form is suggested by theoretical models of
line-driven winds and observations of BAL QSOs (see the
derivation leading to Eqs.~[29] to [31] in G04). The Eddington
luminosity $L_{{\rm Edd}, 46}$, in units of $10^{46}$ erg
s$^{-1}$ is a convenient measure of the BH mass, and
$\epsilon_{\rm QSO}\sim 1-5$ is a strength parameter.

As a consequence, the QSO feedback grows exponentially during
the early phases of galaxy evolution, following the exponential
growth of the supermassive BH mass. It is is negligible in the
first $0.5$ Gyr in all halos, but abruptly becomes notably
important in DM halos more massive than $10^{12}\, M_{\odot}$,
structures weakly affected by SN feedback. Eventually, in these
systems most of the gas becomes unbound from the potential well
of the galaxy halo (see Lapi, Cavaliere \& Menci 2005 for the
impact of QSO feedback on galaxy groups and clusters), so that
star formation and BH activity itself comes to an end on a
timescale which is shorter for more massive galaxies.

Indeed, the positive feedback on BH growth caused by star
formation, in cooperation with the immediate and negative
feedback of SN, and the abrupt and dramatic effect of QSO
feedback, are able to reverse the formation sequence of the
baryonic component of galaxies compared to that of DM halos:
the star formation and the buildup of central BHs are completed
more rapidly in the more massive halos, thus accounting for the
phenomenon now commonly referred to as downsizing.

Before QSO feedback dominates the evolution, radiation is
highly obscured by the surrounding dust. In fact, these
protogalaxies are extremely faint in the UV-optical rest frame
and are more easily selected at submm wavelengths. The nuclear
emission is also heavily obscured, and easier to detect in the
hard X-ray band. On the other hand, when the central
supermassive BH is massive and powerful enough to remove most
of the gas and dust from the surroundings, the active nucleus
shines as an optical QSO. Following this stage, the BH is
already present at the galaxy center, thus any subsequent
supply of gas to the spheroid produces an immediate QSO
feedback, and thus is unable to substantially affect the
stellar or BH mass: afterwards, the stellar populations in the
spheroid evolve largely in a passive manner.

Other SAMs (e.g., Bower et al. 2006; Croton et al. 2006)
introduced the `radiomode feedback', which is active only in
massive objects and at late times to halt cooling flows, but
has no effect during the principal growth phase of most
galaxies and AGNs. Also the highly idealized pseudo AGN
feedback considered by Hatton et al. (2003) and Cattaneo et al.
(2007) \footnote{They simply stops cooling when $\sum M_{\rm
bulge} > 10^{11} M_\odot$, where the sum is over all the
galaxies in a halo.} is somewhat representative of radiomode
feedback. By converse, in G04 a central role is given to the
possible feedback originated by the main episode of
supermassive BH growth, which is responsible for the QSO
activity at high $z$.

The model described above has proved to be extremely successful
in reproducing a wealth of observations, including statistics
of submm galaxies, properties of local ellipticals, the results
of deep $K$-band surveys, demography of supermassive BH relics,
and statistics of high redshift QSOs. These successes are
essentially inherited by its generalization presented here; in
fact, we keep the model parameters fixed to the values used in
the papers by Lapi et al. (2006) and Mao et al. (2007). We list
the model parameters and their fiducial values in Table 1,
stressing their relative relevance in the present context.

\begin{figure}
\centering
\includegraphics[type=eps,ext=.eps,read=.eps, height=8.5cm, angle=0]{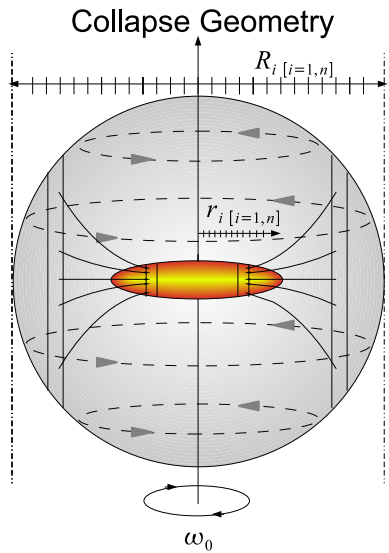}
\caption{A schematic of the disc formation geometry.}
\end{figure}

\begin{table*}
\centering
\caption{Model Parameters}
\begin{tabular}{lcccc}
\hline
\hline
Description &  Symbol & Fiducial value & Reference in the text & Impact on this work\\
\hline
\textbf{Spheroid $+$ BH (\textsl{ABC})}\\
Clumping factor                 &   $\mathcal{C}$            &  $7$    & Eq.~(13) & Strong\\
Radiation drag efficiency       &   $\alpha_{\rm RD}$        &  $2.5$  & Eq.~(16) & Mild\\
Normalization of optical depth  &   $\tau_0$                 &  $1$    & Eq.~(17) & Weak\\
BH radiative efficiency         &   $\eta$                   &  $0.15$ & Eq.~(18) & Mild\\
Eddington ratio                 &   $\lambda_{\rm Edd}$                &  $1$    & Eq.~(18) & Weak\\
SN feedback efficiency          &   $\epsilon_{\rm SN}$      &  $0.05$ & Eq.~(19) & Strong\\
QSO feedback efficiency         &   $\epsilon_{\rm QSO}$     &  $1.3$  & Eq.~(20) & Strong\\
\\
\textbf{Disc (vdB01)}\\
Star formation efficiency         &   $\epsilon_{\rm sf}$    &  $2.5\times 10^{-4}$    & Eq.~(24) & Strong\\
Schmidt law exponent              &   $n$                    &  $1.4$                  & Eq.~(24) & Mild\\
Gas velocity dispersion in Toomre &   $\sigma_{\rm gas}$     &  $6$ km s$^{-1}$        & Eq.~(26) & Weak\\
Normalization constant in Toomre  &   $Q$                    &  $1.5$                  & Eq.~(26) & Weak\\
SN feedback efficiency            &   $\epsilon_{\rm SN}$    &  $10^{-4}$              & Eq.~(27) & Strong\\
\\
\textbf{Dust (\textsl{GRASIL})}\\
Fraction of gas in molecular clouds                  &   $f_{\rm MC}$     &  $0.25$     & Sect. 4.3  & Mild\\
Optical depth of molecular clouds (at $1\,\mu$m)     &   $\tau_{\rm MC}$  &  $60$       & Sect. 4.3  & Weak\\
Escape time from molecular clouds                    &   $t_{\rm esc}$    &  $2.0$ Gyr  & Sect. 4.3  & Weak\\
\hline
\end{tabular}
\hbox{Note. - A Romano IMF $\phi(m_{\star})$ is adopted:
$\phi(m_{\star})\propto m_{\star}^{-1.25}$ for $m_{\star}\geq
M_{\odot}$ and $\phi(m_{\star})\propto m_{\star}^{-0.4}$ for
$m_{\star}\leq M_{\odot}$.}
\end{table*}

\subsection{Modeling the Disc}

At $z<z_t$ during the slow accretion phase, conditions become
sufficiently quiescent to allow the dissipationless growth of
discs from accreting material. Depending on $z_t$ and the shape
of the individual growth history, under suitable circumstances
a substantial disc component may develop. To describe the
process, we adopt the model by van den Bosch (2001 [vdB01]);
here we provide a quick overview of it, but defer the reader to
the original paper for its full description.

At each time step new baryons are accreted onto the halo at a
rate $f_{b}\, \dot{M}_{z}$ proportional to the DM's, in terms
of the universal baryon to DM fraction $f_{b}\approx 0.17$. As
this material enters the halo it is assumed to be heated to the
virial temperature, and to be distributed with an isothermal
profile. The angular momentum distribution of the hot gas
mirrors that of the DM component, so that the change in the
angular momentum over a time interval $\delta t$ reads
\begin{equation}
\delta J=J(t) - J(t-\delta t)~,
\end{equation}
where the halo total angular momentum
$J=G\,M^{5/2}\,\lambda/|E|^{1/2}$ is specified in terms of the
halo spin parameter $\lambda$, see \S~2.

Equating the gained angular momentum to that of a uniformly
rotating shell of material, one obtains the circular frequency
(see also Fig.~9):
\begin{equation}
\omega_{0} = \frac{3}{8\pi}\delta J\, \left( \int~ r^{4}\, \rho(r)\,
\mathrm{d}r\right)^{-1}~.
\end{equation}
The gas is then allowed to cool and collapse, conserving the
initial angular momentum gained from the DM halo. The timescale
for condensation $t_{\rm coll}=\max(t_{\rm dyn}, t_{\rm cool})$
is given by the maximum of the dynamical and cooling time.
After a time $t'=t+t_{\rm coll}$, the cooled gas is added to
the disc annuli with radius $r_i$ corresponding to where it
becomes centrifugally supported upon dissipationless collapse
from the original cylindrical shell radius $R_i$, i.e.,
\begin{equation}
R_i = \left[\frac{r_i\,V_{c}(r_i,t')}{\omega_{0}}\right]^{1/2}~,
\end{equation}
in terms of the local circular velocity $V_c(r)\equiv
[G\,M(<r)/r]^{1/2}$. Thus the disc is allowed to grow in an
onion-like fashion, and in this computation no specific disc
profile is adopted \textit{a priori}.

The star formation rate is then assumed to follow the empirical
Schmidt (1959) law, i.e., it is related to the surface density
$\Sigma(r,t)$ of cold gas in the disc through:
\begin{equation}
\psi(r,t) = \epsilon_{\rm sf}\, \left[{\Sigma(r,t)\over M_{\odot}~
\mathrm{kpc}^{-2}}\right]^{n}~M_{\odot}~\mathrm{kpc}^{-2}~\mathrm{yr}^{-1}~;
\end{equation}
here $\epsilon_{\rm sf}\sim 2.5\times 10^{-4}$ is a fudge
parameter controlling the star formation efficiency, and $n\sim
1.4$ is fixed to match the properties of local spiral galaxies
(Kennicutt 1998).

At each time step and for each annulus within the disc, we
compute the amount of material converted from gas to stars by
solving
\begin{equation}
\frac{\mathrm{d}\Sigma(r,t)}{\mathrm{d}t}=-\psi(r,t)~;
\end{equation}
actually we also impose that a gaseous disc annuli becomes
eligible for star formation only once its surface densities
surpasses a critical threshold $\Sigma_{\rm crit}$ given by the
Toomre (1964) criterion:
\begin{equation}
\Sigma_{\rm crit}={\sigma_{\rm gas}\,\kappa(R)\over 3.36\, G\, Q}~,
\end{equation}
where $\kappa(R)$ is the epicycle frequency (see vdB01 for
details), $Q\sim 1.5$ is a constant, and $\sigma_{\rm gas}\sim
6$ km s$^{-1}$ is the velocity dispersion of the gas.

Cool gas may be removed from the disc through SN winds; we
compute the related mass depletion in a way analogous to the
spheroidal modeling:
\begin{equation}
\dot{M}_{\rm SN}(r,t) = \frac{2 \epsilon_{\rm SN}\,\eta_{\rm SN}
\,E_{\rm SN}}{V_{\rm esc}^{2}(r,t)}\, \psi (r,t)~;
\end{equation}
here in the denominator the local escape velocity $V_{\rm esc}$
is used. Finally, we model the chemical evolution of the
stellar material on using the instantaneous recycling
approximation.

For the sake of coherence, and at variance with vdB01, in the
disc modeling we adopt the same IMF used in the treatment of
the spheroid evolution. We find that the results concerning the
disc structure are affected by $10\%$ from other reasonable
choices of the IMF; we may recover reliable matches to the
properties of the local galaxy population with all commonly
used IMF by altering the fudge parameters of Table 1 within
their physical limits.

The main differences in our modeling with respect to vdB01 are
the following: we use individual growth history derived from
our detailed mergertree, while vdB01 adopts only an averaged
smooth fit; we use a the prescription by Macci\`o et al. (2007)
for the concentration parameter of DM halos, while vdB01 relies
on Bullock et al. (2001); finally, at variance with vdB01 we
take into account the gravitational effect of the preexisting
spheroid on the dynamics of the forming disc but neglect to
model the adiabatic response of the DM halos to disc settling.

\begin{figure}
\includegraphics[type=eps,ext=.eps,read=.eps, height=8.5cm, angle=90]{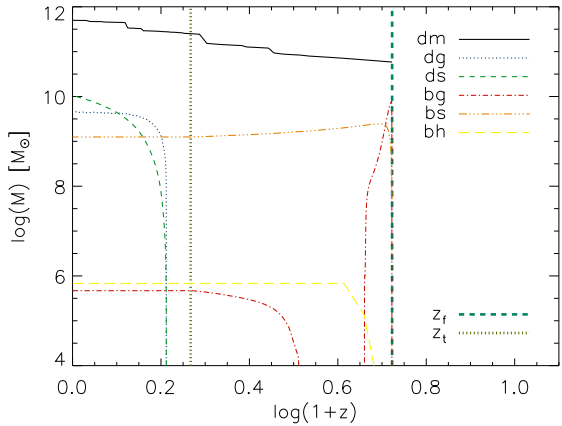}
\caption{Redshift evolution of the baryonic mass components for
a Milky Way sized galaxy at $z=0$: DM mass (solid/black), disc
gas (dotted/blue), disc stars (short-dashed/green), bulge gas
(dot-dashed/red), bulge stars (long-dashed/orange). The two
vertical lines mark the transition and formation redshifts.}
\end{figure}

\begin{figure}
\centering
\includegraphics[type=eps,ext=.eps,read=.eps, height=8.5cm, angle=90]{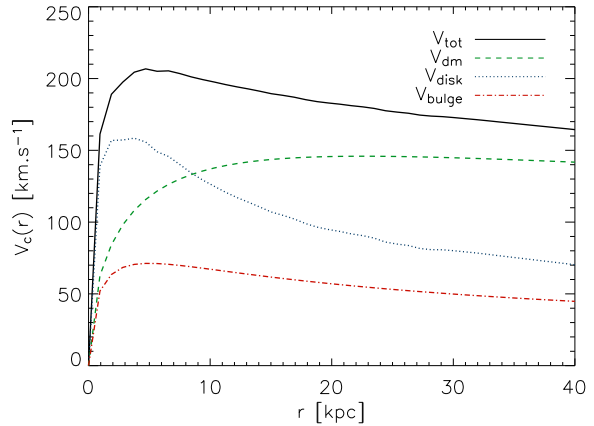}
\caption{Decomposition of the overall rotation curve (solid
line) for our Milky Way-type galaxy at $z=0$ in terms of the
contributions from DM (dashed/green line), bulge
(dot-dashed/blue line) and disc (dotted/blue line).}
\end{figure}

\begin{figure}\centering
\includegraphics[type=eps,ext=.eps,read=.eps, height=8.5cm, angle=90]{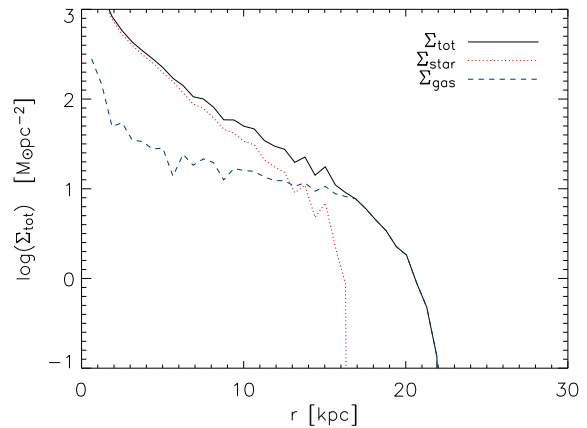}
\caption{Decomposition of the disc surface density profiles
(solid line) for a Milky Way sized galaxy at $z=0$, in terms of
the stellar (dotted-line) and gaseous (dashed line)
components.}
\end{figure}

\begin{figure}
\centering
\includegraphics[type=eps,ext=.eps,read=.eps, height=8.5cm, angle=90]{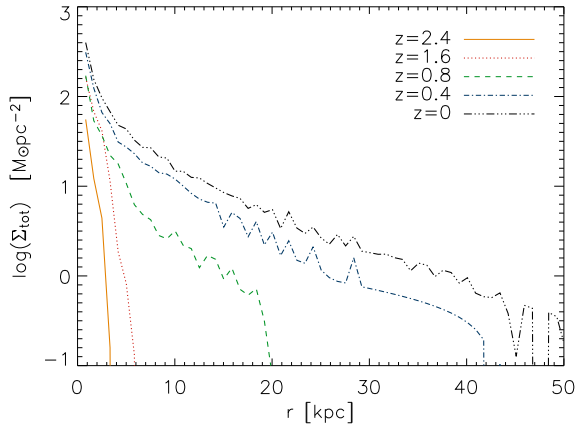}
\caption{Redshift evolution of the overall surface density
profile for a $L_{\star}$ galaxy with spin parameter
$\lambda=0.06$ resulting in an extension that is twice the
fiducial model to highlight the evolution. Note that at
$z\approx 0.8$ a minor merger event occurs, the exponential
disc structure is temporarily disrupted, and an anti-truncated
disc (see text for details) develops.}
\end{figure}

\begin{figure}
\centering
\includegraphics[type=eps,ext=.eps,read=.eps, height=8.5cm, angle=90]{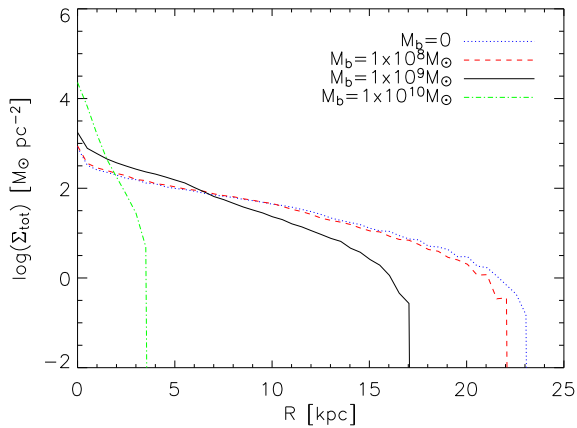}
\caption{Impact of preexisting bulge masses on the disc surface
density profiles (solid line) for a Milky Way-sized galaxy
halo $M_0\approx 5\times 10^{11}\, M_{\odot}$ at $z=0$; as the
bulge mass approaches the disc mass of about
$10^{10}\,M_{\odot}$ discs become more compact. Higher
bulge masses yield significantly more concentrated discs.}
\end{figure}

\section{Results}

\subsection{Disc structure and dynamics}

In this Section we analyze the behavior of individual galaxies,
focusing on the disc component; the formation and evolution of
the spheroidal component has been extensively considered in
several previous papers by our team (G04, Silva et al. 2005,
Cirasuolo et al. 2005, Lapi et al. 2006).

To this purpose we select a fiducial model galaxy with current
mass $M_0\approx 5\times 10^{11}\,M_{\odot}$ similar to that of
the Milky Way (see Xue et al. 2008; Naab \& Ostriker 2006), and
with an average spin parameter $\lambda = 0.04$ (see Macci\`o
et al. 2008). We find that the resulting galaxy components at
$z=0$ are in generally good agreement with observed Milky Way
properties, finding that $M_{\rm disc,0}\approx 1.5 \times
10^{10}\, M_{\odot}$, $M_{\rm bulge,0}\approx 1.3 \times
10^{9}\,M_{\odot}$, see also Naab \& Ostriker (2006).

In Fig.~10 one can see the buildup of the various components
for this fiducial galaxy. At $z > z_t$ a strong growth of the
spheroidal component takes place which is halted by the QSO
activity after approximately $10^8$ years from $z_f$. Following
this, the stellar populations in the spheroid evolve passively,
and the residual gaseous material is originated from the
stellar recycling. At $z < z_t$ new gas quiescently accretes
onto the halo forming a disc structure; note that star
formation is delayed until the cold gas surface density becomes
sufficiently large to overcome the critical star formation
threshold.

In Fig.~11 we present the rotation curve decomposition for our
fiducial galaxy. The total rotation curve is flat out to large
radii; there it is DM dominated, while in the inner regions the
baryonic components of the disc and the bulge dominate the
gravitational potential, in agreement with kinematic models of
the Milky Way; we find a peak rotation velocity $V_{\rm
max}\approx 210$ km s$^{-1}$.

Fig.~12 shows the disc radial surface density profile at $z=0$;
we obtain general exponential stellar profile out to a
truncation radius of $16.2$ kpc, in broad agreement with the
value of $12$ kpc observed for the Milky Way (see Naab \&
Ostriker 2006). The gaseous disc is more extended than the
stellar one due to the critical star formation threshold.
However, note that the gaseous disc is depleted in the central
regions because of star formation, and there the stellar
component dominates. Fitting an exponential profile
$\Sigma_{0}\,e^{-r/r_d}$ to the stellar disc, we find an
exponential scale radius $r_d\approx 2.6$ kpc, which is
comparable to observational estimates $2.5-3.5$ kpc for the
Milky Way, see Sackett (1997).  Finally, within this model, we
find a BH mass at $z=0$ of $M_{\rm BH,0}\approx 6.7 \times
10^{5}\, M_{\odot}$, which is less massive but still consistent
with the one at the center of the Milky Way $M_{\rm
BH,0}\approx 3.6 \times 10^{6}\, M_{\odot}$, see Eisenhauer et
al. (2005).

For the sake of completeness we highlight the related redshift
evolution of the disc profile in Fig.~13. For illustrative
purposes it is clearer to have an elongated disc profile; thus
we choose a realization with a spin parameter $\lambda=0.06$
larger than the fiducial value. The disc (gas and stars)
naturally evolves from the inside-out, retaining a quasi
exponential profile with scalelength increasing over time.
Although the vast majority of model discs have a
quasi-exponential surface density profiles, the detailed shape
depends on the details of the specific growth history, and in
particular on the transition redshift $z_t$.

We stress that our discs develop following the buildup of the
spheroidal inner component, which affects the overall
gravitational potential. Fig.~14 illustrates how the $z=0$ disc
surface density depends on the mass of a preexisting bulge. We
find that for identical realizations for our fiducial model,
but imparting a bulge mass by hand, the disc structure (and
thus its evolution) is altered. This dynamical interdependence
results in a disc structure which becomes significantly more
compact as the bulge mass approaches the disc mass at $z=0$. In
addition, we find that with the presence of a substantial
spheroid component, higher transition redshifts $z_t$ yield
more extended discs at $z=0$.

Recent observational studies have shown that in general
exponential discs come in three categories, corresponding to
simple exponential (type I), truncated (type II) and
anti-truncated (type III) surface density profiles (Pohlen \&
Trujillo 2006). We find that exponential surface density
profiles (of type I and II) are generated as a generic feature
of our model (see Fig.~12). Interestingly, discs which have
undergone a recent minor merger event have a significant amount
of material added to the outer parts in a non smooth fashion,
and these mimic disc anti-truncations (type III, see Fig.~13 at
$z\approx 0.8$). However, the disc anti-truncations are
typically masked through subsequent accretion, that restore the
overall exponential surface densities. We plan to address this
issue in a subsequent work.

\subsection{Galaxy properties at $z=0$}

Now we turn to study the properties of the local galaxy
population; to this purpose we generate catalogues of galaxies
that encompass a representative range of $z=0$ halo masses,
from $10^9$ to $10^{14}\, M_{\odot}$ in logarithmic increments.
We then exploit the statistics of halos containing one single
galaxy, namely, the galaxy halo mass function (GHMF) as
provided by Shankar et al. (2006); the latter authors provide
the following analytic fit
\begin{equation}
\Theta(M_0)=\frac{\theta}{\bar{M}}\left(\frac{M_0}{\bar{M}}\right)^{\alpha}\,
e^{-M_0/\bar{M}}~,
\end{equation}
with $\alpha\approx 1.84$, $\theta\approx 3.1\times 10^{-4}$
Mpc$^{-3}$, and $\bar{M}\approx 1.12\times 10^{13}\,M_{\odot}$.
This function is derived by subtracting the group and cluster
mass function (Martinez et al. 2002) from the Sheth \& Tormen
(1999) mass function. Thus, for halos with $M_0 <$ a few
$10^{13}\, M_{\odot}$ the GHMF closely follow the Sheth \&
Tormen mass function, while the falloff at larger masses
mirrors the increasing probability of multiple occupation. In
principle, we may account for the halo occupation number
through the full mergertree, but this will introduce
uncertainties related to poor knowledge of processes like
dynamical friction, tidal stripping etc. etc.; thus we prefer
to bypass this problem using the GHMF.

\begin{figure}
\centering
\includegraphics[type=eps,ext=.eps,read=.eps, height=8.5cm, angle=90]{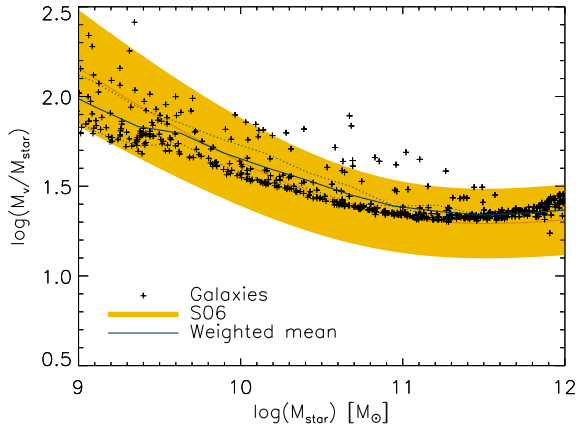}
\caption{The relation between the host DM halo mass and the
stellar mass ratio. Model results (over $1000$ realizations)
for individual galaxies (crosses), model average (solid line)
and quartiles (dashed lines) are compared to the observational
determination by Shankar et al. (2006, shaded area).}
\end{figure}

\begin{figure}
\centering
\includegraphics[type=eps,ext=.eps,read=.eps, height=8.5cm, angle=90]{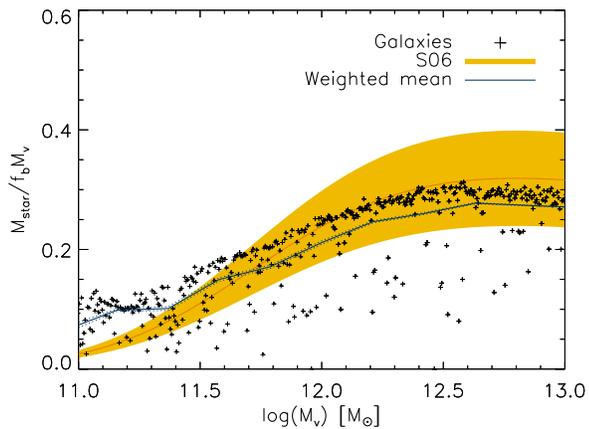}
\caption{Fraction of baryons converted into stars as a function
of host halo mass. Model results (over $1000$ realizations) for
individual galaxies (crosses), model average (solid line) and
quartiles (dashed lines) are compared to the observational
determination by Shankar et al. (2006, shaded area).}
\end{figure}

\begin{figure}
\centering
\includegraphics[type=eps,ext=.eps,read=.eps, height=8.5cm,angle=90]{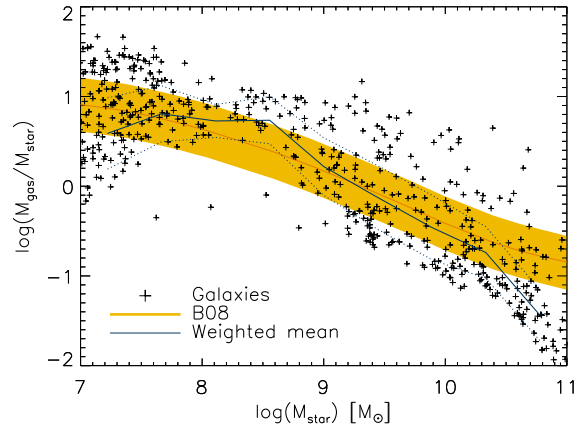}
\caption{Gas to stellar mass fraction as a function of the
stellar mass. Model results (over $1000$ realizations) for
individual galaxies (crosses), model average (solid line) and
quartiles (dashed lines) are compared to the observational
determination by Baldry et al. (2008, shaded area).}
\end{figure}

\begin{figure}
\centering
\includegraphics[type=eps,ext=.eps,read=.eps, height=8.5cm, angle=90]{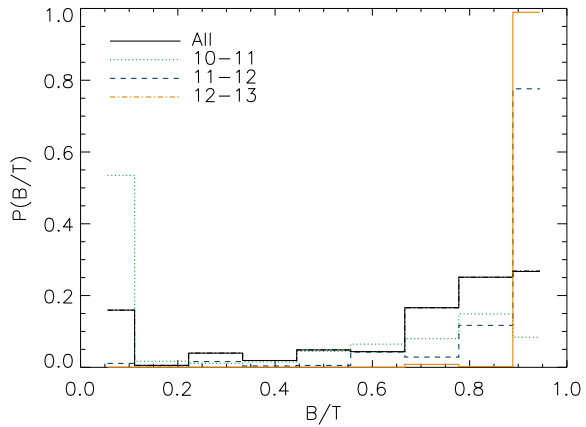}
\caption{The occurrence of bulge to total mass ratio in our
model, binned for different halo mass ranges (results over
$1000$ realizations); disc-dominated galaxies occur
preferentially in low-mass halos, while spheroid-dominated
galaxies occur preferentially in massive halos.}
\end{figure}

In Figs.~14-16 we compare our model predictions with the
results of Shankar et al. (2006) and Baldry et al. (2008), who
derive a number of galaxy properties as a function of the host
DM halo mass. In Fig.~15 we consider the fraction of stellar to
total mass within DM halos; we see that there is a steep
increase in the DM dominance for low mass halos, since these
provide inefficient environments for star formation mostly due
to the impact of SN feedback.

Fig.~16, constituting a different rendition of the previous
plot, directly highlights the fraction of available baryons
condensed into stars as a function of the host halo mass. In
halos of masses exceeding few $10^{12}\, M_{\odot}$ star
formation is more efficient; in the absence of a substantial
impact of QSO feedback the efficiency would keep growing with
increasing mass (despite an increasing difficulty of the
cooling processes), while both the data and our model show a
clear flattening.

Fig.~17 illustrates the correlation between gas fraction (ratio
of the total cold gas to the total baryonic mass within the
galaxy) and the overall stellar mass, compared to the data by
Baldry et al. (2008); less massive galaxies typically have a
significantly larger gas fraction. This is because more massive
galaxies are typically spheroid dominated and thus underwent
strong gas ejection by the QSO feedback. On the other hand,
lower mass galaxies are typically disc dominated, QSO feedback
thus is relatively unimportant, and also the critical surface
density threshold becomes increasingly difficult to surpass.

Fig.~18 illustrates the occurrence of bulge to total mass ratio
in our model, binned in host halo mass. The behavior of the
galaxies in our model is dichotomic, with the disc-dominated
galaxies to occur preferentially in low-mass halos, while
spheroid-dominated galaxies to occur preferentially in massive
halos. This result is basically linked to the distribution of
transition redshift $z_t$ discussed in \S~2 and illustrated in
Fig.~6.

Although not reported here, we stress again that this model
inherits from the \textsl{ABC} scenario the good match with the
observed local BH mass vs. bulge relationships (see G04,
Cirasuolo et al. 2005, Lapi et al. 2006).

\begin{figure}
\centering
\includegraphics[type=eps,ext=.eps,read=.eps, height=8.5cm, angle=90]{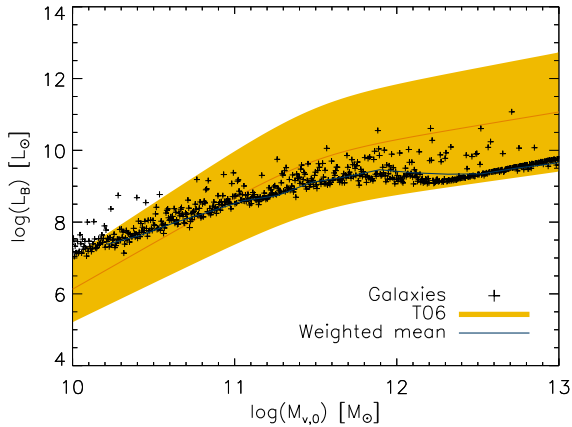}
\caption{$B$-band luminosity as a function of the host DM halo
mass. Model results (over $1000$ realizations) for individual
galaxies (crosses), model average (solid line) and quartiles
(dashed lines) are compared to the observational determinations
collected by Tonini et al. (2006, shaded area).}
\end{figure}

\begin{figure}
\centering
\includegraphics[type=eps,ext=.eps,read=.eps, height=8.5cm, angle=90]{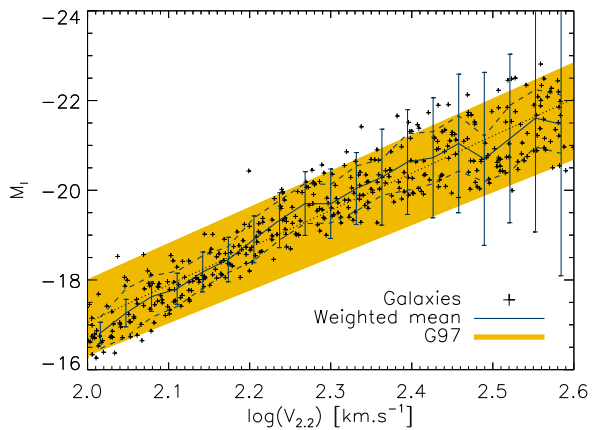}
\caption{The Tully-Fisher relation. Model results (over $1000$
realizations) for individual galaxies (crosses), model average
(solid line) and quartiles (dashed lines) are compared to the
observational determination by Giovanelli et al. (1997, shaded
area). The Possinian error bars illustrate the relative
abundance of galaxies with different circular velocity.}
\end{figure}

\subsection{Spectrophotometric properties}

In order to analyze the luminous properties of galaxies, we
interface our model with the spectrophotometric code
\textsl{GRASIL} (Silva et al. 1998), that accounts for the
attenuation and reradiation of starlight by dust.

\textsl{GRASIL} uses stellar population synthesis models based
on the Padova evolutionary tracks, which include the
effects of dusty envelopes around asymptotic giant branch stars
(Bressan, Granato \& Silva 1998). Then each single stellar
population is summed taking into account the appropriate age
and metallicity, and weighted with the star formation rate to
obtain the unattenuated SED
\begin{equation}
F_{\lambda}(\tau)= \int_{0}^{\tau}
\zeta_{\lambda}[\tau-t,Z(t)]\,\psi(t)\, \mathrm{d}t~;
\end{equation}
here $\tau$ is the age of the galaxy, $t$ is the birth age of
an individual single stellar population $\zeta_{\lambda}(t,Z)$,
and $\psi(t)$ is the star formation rate.

For the detailed description of dust attenuation and
reprocessing of starlight we defer the reader to the papers by
Silva et al. (1998, 2005). We just stress here that
\textsl{GRASIL} includes the effect of differential dust
extinction of stellar population, i.e., younger stellar
generation are more affected by dust obscuration; this is
because stars form in molecular clouds, an environment denser
than the average, and progressively get rid of them.

The \textsl{GRASIL} SEDs depend on the following basic
parameters: the fraction $f_{\rm MC}$ of gas in the form of
molecular clouds rather than in the diffuse interstellar
medium; the optical depth $\tau_{\rm MC}$ of molecular clouds
to the radiation emitted from a source at their center (at
$1\,\mu$m); the escape time $t_{\rm esc}$ of newly born stars
from molecular clouds. On the basis of previous works in which
\textsl{GRASIL} has been coupled with various SAMs, and in
particular with the one of the Durham team (see Granato et al.
2000 for details; also Baugh et al. 2005, Monaco et al. 2007),
we set the \textsl{GRASIL} parameters to the standard values
reported by Silva et al. (2005) and listed in Table 1. Note
that since in this paper we do not consider regions of the SED
strongly affected by dust emission, the dependence of our
results on these parameters is mild/weak. Due to the
preliminary nature of this work, we do not exploit the
multiwavelength capabilities of \textsl{GRASIL} to the full
extent, but focus on reproducing several local galaxy
population properties in selected bands; we delay a more
refined analysis for future work.

In Fig.~19 we show the $B$-band luminosity of our model
galaxies as a function of DM halo mass. We find a relatively
strong correlation with little scatter, in general agreement
with the data by Tonini et al. (2006). The break around
$10^{11}\, M_{\odot}$ is due to the impact of SN feedback in
small systems, where star formation becomes progressively less
efficient.

Fig.~20 illustrates the I-band Tully-Fisher relation from our
model; note that we extract the peak rotation velocity from our
model by fitting an exponential to the disc surface density
profile. Then we use the obtained scale radius $r_d$ to define
the maximum velocity as $V_{\rm max} = V_c(2.2\,r_d)$ in
analogy with the observational methods; this procedure
therefore does not resort to further assumptions about disc
structure and dynamics.

Our model result is compared with the data by Giovanelli et al.
(1997), finding excellent agreement in both slope and
normalization; fitting our result with the law $M_I = m\,
[\log_{10}(V_{2.2})-2.5]+ c$, we obtain $m=-8.30$, and
$c=-21.19$, to be compared with the observational values
$m=-7.68$ and $c=-21.0$. We also represent the intrinsic
scatter in our results by the blue dashed contours, showing a
slight if systematic increase in scatter towards higher
rotational velocities, in general agreement with observations
(Giovanelli et al. 1997). Finally, we show the relative
abundances of galaxies as represented by the Poissonian error
bars that account for the relative numbers of halos within a
cosmological volume. We see that the slowly rotating, faint
galaxies are more abundant; galaxies become rarer as we move
toward larger circular velocities, and within our sample we do
not find any galaxies with rotational velocities larger than
$V_{2.2} > 10^{2.6}$ km s$^{-1}$.

In Fig.~21 we present the $r^{\star}$-band luminosity function from
our model at $z=0$, and compare it to the fit by Benson et al.
(2007) based on SDSS data; we find an overall good agreement.
We confirm, in tune with a number of previous works, that the
flattening at the faint end is mainly due to the impact of SN
feedback in small systems (Benson et al. 2003 and references
therein), and that the steepening at the bright end is mainly
caused by the impact of QSO feedback in massive galaxies
(Somerville et al. 2008 and references therein). In addition,
we highlight the different contributions to the overall
luminosity function from the spheroid and the disc component;
the latter typically dominate the faint end, while the former
dominate the bright end, as expected on an intuitive basis (see
also Kauffmann, White \& Guiderdoni 1993; Baugh, Cole \& Frenk
1996; Kauffmann \& Charlot 1998; Somerville \& Primack 1999;
Hatton et al. 2003; Tasca \& White 2005).

We stress that the simultaneous fitting of the galaxy
luminosity function and of Tully-Fisher relation is challenging
for many SAMs, and so constitutes a big success of our model
(see Courteau et al. 2007; Dutton et al. 2007; Bell et al.
2003).

\begin{figure}
\centering
\includegraphics[type=eps,ext=.eps,read=.eps, height=17.cm,  angle=90, clip=1]{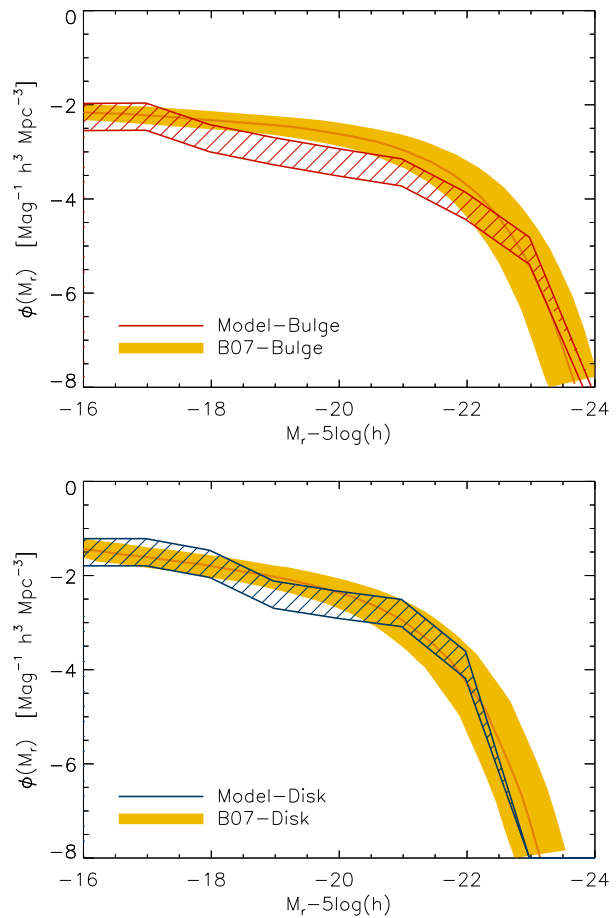}\\
\caption{The $r^*-$band luminosity function. The overall
results (over $1000$ realizations), with the contributions from
spheroids (red hatched region - upper panel) and discs (blue
hatched region - lower panel) highlighted, is compared to the
fits derived from the SDSS data by Benson et al. (2007,
coloured dashed lines).}
\end{figure}

\section{Discussion and Conclusions}

In this paper we have proposed a novel scenario for the
formation and evolution of galaxies in the standard
$\Lambda$CDM framework.

We have been motivated by several recent high-resolution
$N$-body simulations (Zhao et al. 2003; Diemand et al. 2007;
Hoffmann et al. 2007; Ascasibar \& Gottloeber 2008), that
recognize the DM halo growth to occur in two rather distinct
phases: an early violent collapse featuring a few major
mergers, and a late quiescent accretion onto the halo outskirts
that does not affect the inner regions where the galactic
structure resides. We associate these two phases to two
different modes of galaxy formation, leading to spheroids and
discs.

Specifically, we envisage that spheroids form during the fast
collapse phase, when violent major mergers reshuffle the
gravitational potential and cause dynamical relaxation and
orbit isotropization of the DM and stellar components (see Lapi
\& Cavaliere 2009). Meanwhile, strong starburst activity and
the growth of a central supermassive BH take place in parallel.
The ensuing SN explosions and the nuclear activity feed energy
back to the baryons, and regulate the ongoing star formation
rate and BH growth. These mutual energy feedbacks actually
reverse the formation sequence of the baryonic component of
galaxies compared to that of DM haloes: the star formation and
the buildup of central BHs are completed more rapidly in the
more massive haloes, thus accounting for the phenomenon now
commonly referred to as downsizing. In the subsequent slow
accretion phase, during which major mergers are rare, the
quiescent growth of a disclike structure around the preformed
spheroids can occur by dissipantionless collapse.

We then test this new scenario against observations resorting
to the semianalytic technique. To this purpose, we adopt
standard and widespread models. As to the DM evolution we base
on the algorithm by Cole et al. (2000) and Parkinson et al.
(2008) supplemented by the results of the $N$-body experiments
by Zhao et al. (2003). As to the spheroid component we rely on
the prescriptions by G04 and following developments. As to the
disc buildup we base on the recipes by vdB01. Finally, we
couple everything to the spectrophotometric code
\textsl{GRASIL} by Silva et al. (1998).

Note that the SAM developed here can be viewed as an extension
of our previously proposed \textsl{ABC} model (see G04). The
latter dealt with the high-redshift spheroid-supermassive BH
formation, and proven to be successfull in many respects (see
Silva et al. 2005, Cirasuolo et al. 2005, Lapi et al. 2006, Mao
et al. 2007, Lapi et al. 2008). Practically, we now include the
disc formation at low redshift, so extending it to encompass
all morphological galaxy types and cosmic epochs.

Though we are confident to have described through conceivable
physical recipes the key processes ruling galaxy formation and
evolution, we must admit that our modeling disregard or treat
crudely several aspects that may play a relevant role: baryon
impacts on the detailed structure of DM halos, environmental
effects, angular momentum evolution, bar instabilities, halo
occupation distributions, etc. In fact, in this preliminary
study we pursue the strategy of `keeping it as simple as
possible', our aim being to test with minimal ingredients
whether our scenario could provide results in accord at least
with the local galactic observables, and eventually it proved
to perform such a remarkable task surprisingly well.

Specifically, we have shown our model to reproduce the observed
stellar mass fractions (see Figs. 15 and 16), gas content (see
Fig. 17), morphological dichotomy (see Fig. 18), mass-to-light
ratios (see Fig. 19), Tully-Fisher relation (see Fig. 20), and
luminosity functions (see Fig. 21) of the local galaxy
populations. In future works we aim to compare our model
predictions to the intermediate- and high-redshift data;
however, note that at $z\ga 2$ our model, built upon the SAM by
G04, still performs quite well by construction. We will also
pursue the analysis of galaxy statistics at multiple
wavelengths; this should allow us to understand better the
interplay between the processes involved within our scenario.
Finally, we will discuss more extensively the structural
properties of the discs emerging from our model, that will
constitute testbeds for the next generation of SAMs.

\section*{Acknowledgments}
We thank an anonymous referee for constructive comments and
helpful suggestions. We acknowledge stimulating discussions
with L. Silva and A. Schurer. MC has been supported through a
Marie Curie studentship for the Sixth Framework Research and
Training Network MAGPOP, contract number MRTN-CT-2004-503929.
AL was supported in part by ASI, and thanks INAF-OATS for kind
hospitality.

\end{document}